\documentclass[final,3p,times]{elsarticle}
\usepackage{amssymb}
\usepackage{amsmath}
\usepackage{subcaption}

\begin{document}

\begin{frontmatter}

%% Title, authors and addresses

%% use the tnoteref command within \title for footnotes;
%% use the tnotetext command for theassociated footnote;
%% use the fnref command within \author or \affiliation for footnotes;
%% use the fntext command for theassociated footnote;
%% use the corref command within \author for corresponding author footnotes;
%% use the cortext command for theassociated footnote;
%% use the ead command for the email address,
%% and the form \ead[url] for the home page:
%% \title{Title\tnoteref{label1}}
%% \tnotetext[label1]{}
%% \author{Name\corref{cor1}\fnref{label2}}
%% \ead{email address}
%% \ead[url]{home page}
%% \fntext[label2]{}
%% \cortext[cor1]{}
%% \affiliation{organization={},
%%             addressline={},
%%             city={},
%%             postcode={},
%%             state={},
%%             country={}}
%% \fntext[label3]{}

\title{sPHENIX measurement of Open-Charm Baryon-to-Meson Ratios in $p$+$p$ collisions at RHIC}

%% use optional labels to link authors explicitly to addresses:
%% \author[label1,label2]{}
%% \affiliation[label1]{organization={},
%%             addressline={},
%%             city={},
%%             postcode={},
%%             state={},
%%             country={}}
%%
%% \affiliation[label2]{organization={},
%%             addressline={},
%%             city={},
%%             postcode={},
%%             state={},
%%             country={}}

\author{Xudong Yu On behalf of the sPHENIX Collaboration} %% Author name
\ead{yuxd@stu.pku.edu.cn}

%% Author affiliation
\affiliation{organization={School of Physics,Peking University},%Department and Organization
            addressline={209 Chengfu Road}, 
            city={Beijing},
            postcode={100871}, 
           % state={},
            country={People’s Republic of China}}

%% Abstract
\begin{abstract}
sPHENIX is a state-of-the-art experiment at the Relativistic Heavy Ion Collider (RHIC), dedicated to the study of heavy-flavor and jet physics. Its precision tracking system, combined with streaming readout, enables heavy-flavor measurements with high-statistics and essentially unbiased data samples. During the 2024 run, sPHENIX was fully commissioned and recorded a sample of 100 billion unbiased $p$+$p$ collisions, together with a minimum-bias Au+Au dataset. The 2025 run further expanded the sPHENIX dataset with high-statistics $p$+$p$, O+O and Au+Au collisions. This extensive $p$+$p$ sample opens the door to heavy-flavor measurements with orders of magnitude more statistics than previously available at RHIC. Notably, there has been no prior measurement of the $\Lambda_c^+ / D^0$ baseline in $p$+$p$ collisions at RHIC energies. The large sPHENIX dataset now enables the first exploration of key open questions, such as the hadronization mechanism of baryons and the strange-to-light flavor meson ratio.
%We will present the status of the first measurements of the $\Lambda_c^+ / D^0$ ratio and the similarly unexplored $D_s^+ / D^+$ ratio in $p$+$p$ collisions.
\end{abstract}

%%Graphical abstract
%\begin{graphicalabstract}
%\includegraphics{grabs}
%\end{graphicalabstract}

%%Research highlights
%\begin{highlights}
%\item Research highlight 1
%\item Research highlight 2
%\end{highlights}

%% Keywords
\begin{keyword}
Open charm\sep Baryon-to-Meson Ratios \sep sPHENIX
%% keywords here, in the form: keyword \sep keyword

%% PACS codes here, in the form: \PACS code \sep code

%% MSC codes here, in the form: \MSC code \sep code
%% or \MSC[2008] code \sep code (2000 is the default)

\end{keyword}

\end{frontmatter}

%% Add \usepackage{lineno} before \begin{document} and uncomment 
%% following line to enable line numbers
%% \linenumbers

%% main text
%%

%% Use \section commands to start a section
\section{Introduction}\label{sec:intro}

Charm quarks are produced predominantly in initial hard scatterings and therefore probe the earliest stage of the collision. Owing to their large masses, they are generated on short time scales and subsequently experience the full space-time evolution of the system. Open-charm hadrons are thus powerful tools for studying both the perturbative production of heavy quarks and the non-perturbative dynamics of hadronization. The latter remains one of the central challenges in QCD. In elementary systems, such as $e^+e^-$ and $e^-p$ collisions, charm-hadron production is often described in terms of fragmentation functions that are commonly assumed to be universal. However, measurements at the Large Hadron Collider (LHC) have shown that the $\Lambda_c^+/D^0$ ratio in $p$+$p$ collisions is significantly larger than expectations based on fragmentation functions extracted from $e^+e^-$ and $e^-p$ data~\cite{CMS:2019uws,ALICE:2020wfu,ALICE:2023sgl}. This observation suggests that the universality of fragmentation may be violated in hadronic collisions and that additional hadronization mechanisms may contribute. Several scenarios have been proposed to explain the enhancement, including color reconnection~\cite{Christiansen:2015yqa}, statistical hadronization~\cite{He:2019tik}, and coalescence~\cite{Plumari:2017ntm}. More precise measurements are therefore essential for discriminating among these mechanisms and constraining theoretical descriptions. Ratios involving strange and non-strange charm mesons provide complementary information. In particular, the $D_s^+$ meson contains a strange quark, whereas the $D^0$ and $D^+$ do not. Measurements of $D_s^+$ production relative to non-strange charm mesons can therefore provide insight into strange-quark abundance and its role in charm-hadron formation. At RHIC energies, such measurements offer a valuable complement to those performed at the LHC by probing a different collision-energy regime.

The STAR experiment has measured the $\Lambda_c^+/D^0$ ratio in Au+Au collisions~\cite{STAR:2019ank}, but no corresponding $p$+$p$ measurement exists at RHIC. Such a baseline is indispensable for interpreting heavy-ion results. The sPHENIX experiment is a state-of-the-art collider detector at the Relativistic Heavy Ion Collider (RHIC) at Brookhaven National Laboratory~\cite{PHENIX:2015siv}. One of its primary physics goals is to study the properties of the quark-gluon plasma (QGP) with hard probes, including jets and heavy-flavor hadrons, especially in the low-$p_T$ region where RHIC measurements are particularly competitive and complementary to those at the LHC.

The construction of sPHENIX was completed in 2023. Commissioning began with Au+Au collisions in 2023 and continued with $p$+$p$ collisions in 2024. After commissioning, sPHENIX collected a large sample of unbiased $p$+$p$ collisions by exploiting its streaming readout capability. All results presented in these proceedings are based on the 2024 $p$+$p$ dataset collected by sPHENIX. We first provide a brief overview of the detector, then summarize the current tracking performance, and finally present the status of the ongoing open-charm analyses.

\section{sPHENIX detector and tracking reconstruction}

The sPHENIX detector comprises a precision tracking system and large-acceptance electromagnetic and hadronic calorimeters in the central barrel, together with forward and far-forward detectors, including the Minimum Bias Detector (MBD), the Event Plane Detector (sEPD), and the Zero-Degree Calorimeter (ZDC). The central barrel covers $|\eta| < 1.1$ with full azimuthal acceptance, while the MBD, sEPD, and ZDC are installed outside this region. Four dedicated tracking subsystems are used for charged-particle reconstruction and heavy-flavor measurements. The Monolithic Active Pixel Sensor (MAPS)-based vertex detector (MVTX) consists of three layers of silicon pixel sensors located close to the beam pipe and provides the primary vertex resolutions less than 10 microns, which are essential for heavy flavor reconstruction. The Intermediate silicon Tracker (INTT) is a two-layer silicon strip detector with timing resolution sufficient to resolve individual 106.5~ns RHIC bunch crossings in $p$+$p$ collisions, enabling proper track-to-crossing association and mitigating pile-up effects. The Time Projection Chamber (TPC) is a gaseous tracker with 48 layers spanning radii from 20 to 78~cm and equipped with GEM-based continuous readout. It delivers the momentum resolution required for the sPHENIX tracking program. Although the TPC is not optimized for particle identification, its $\mathrm{d}E/\mathrm{d}x$ information provides useful separation power at low momentum and is therefore valuable for suppressing combinatorial background in heavy-flavor analyses. Outside the TPC, the TPC Outer Tracker (TPOT), an eight-tile Micromegas detector, provides additional space points and is important for correcting TPC space-charge distortions. Together, these subsystems enable four-dimensional track reconstruction, including both spatial and timing information. Tracks are reconstructed through hit clustering in the individual detectors, seed finding in the silicon and TPC subsystems, space-time matching of track seeds, and a final Kalman-filter fit performed with the \textsc{Acts} software package~\cite{Osborn:2021zlr,Ai:2021ghi}.

\begin{figure}[htbp]
    \centering
    \begin{subfigure}[h]{0.42\linewidth}
    {
        \includegraphics[width=\linewidth]{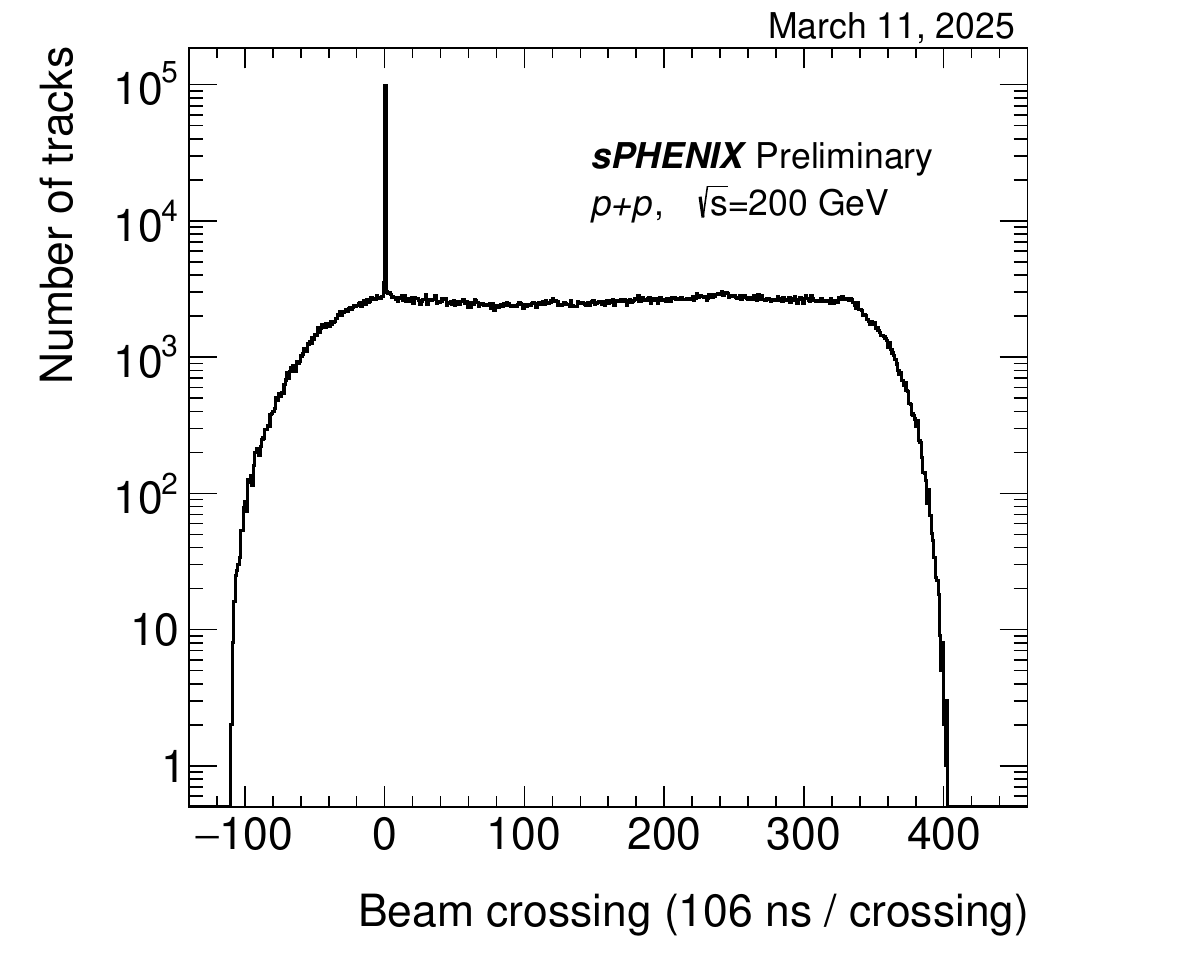}
        \caption{}
        \label{fig:crossing-track}
    }
    \end{subfigure}
    \begin{subfigure}[h]{0.35\linewidth}
    {
        \includegraphics[width=\linewidth]{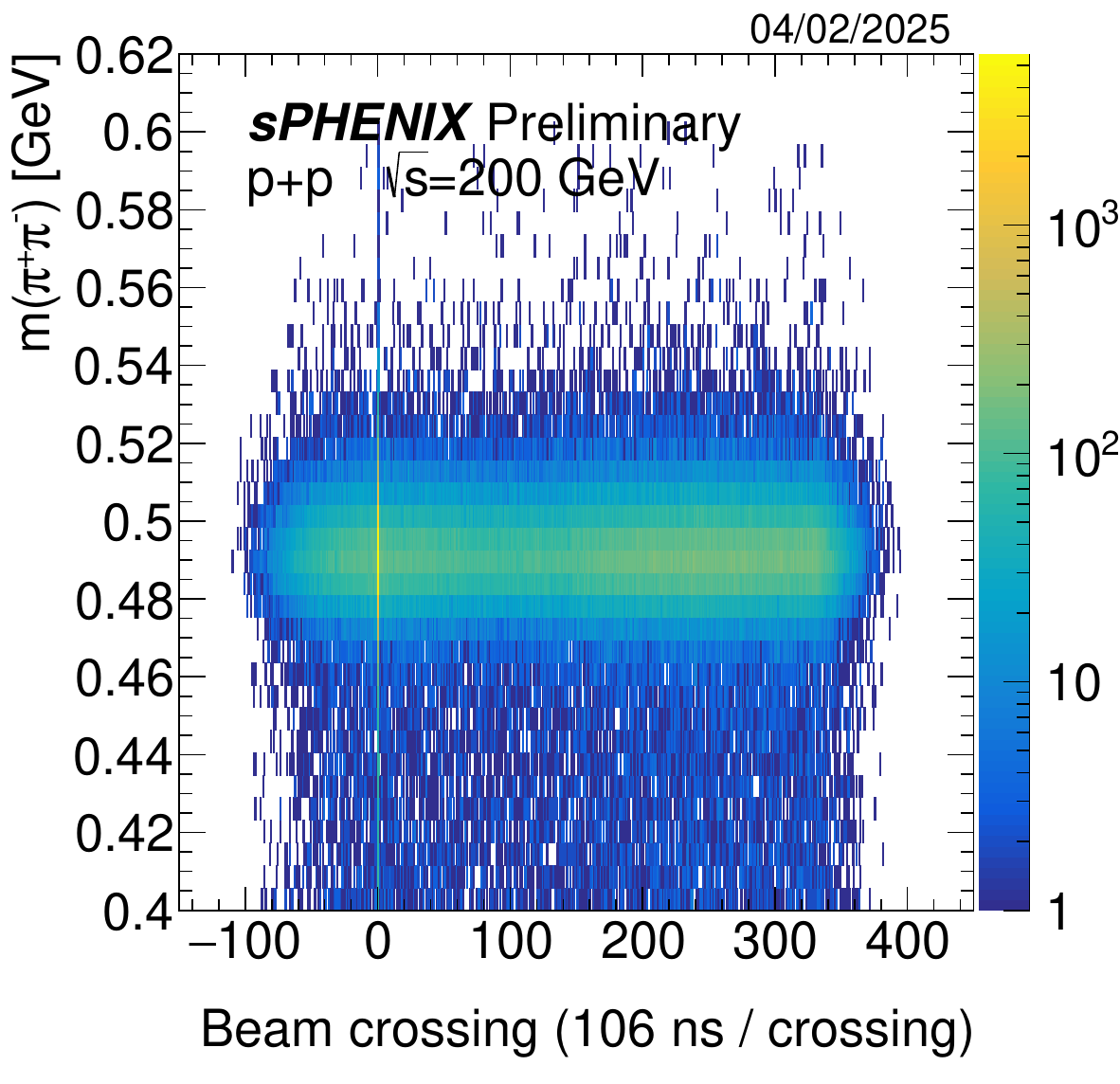}
        \caption{}
        \label{fig:crossing-ks}
    }
    \end{subfigure}
    \caption{(a) The number of reconstructed tracks in $\sqrt{s}=200$ GeV $p$+$p$ collisions as a function of the relative beam crossing number, where crossing 0 corresponds to the hardware trigger timing. (b) Two-dimensional distribution of $\pi^+\pi^-$ invariant mass distribution versus relative beam crossing number.}
    \label{fig:crossing}
\end{figure}

A unique feature of the sPHENIX tracking detectors is their extended streaming readout mode, in which approximately 20\% to 30\% of the $p$+$p$ collisions delivered by RHIC are recorded in Run~24. This mode is particularly important for open-heavy-flavor physics, where low-$p_T$ hadrons suffer from poor hardware-trigger efficiencies. By the end of Run~24, the tracking detectors were recording unbiased $p$+$p$ collisions at a rate of $\mathcal{O}(200)$~kHz, about 20-50 times higher than that achievable with a purely minimum-bias trigger. Figure~\ref{fig:crossing-track} shows the number of reconstructed tracks as a function of the beam-crossing number relative to the hardware trigger, where crossing 0 corresponds to the triggered bunch crossing and the other crossings arise from the streamed data. The approximately constant yield across crossings demonstrates stable tracking performance in streaming mode. Figure~\ref{fig:crossing-ks} shows the two-dimensional distribution of the $\pi^+\pi^-$ invariant mass versus the relative beam-crossing number. The visible $K_S^0$ band across all crossings further confirms the uniform reconstruction performance. Over the full Run~24 period, sPHENIX recorded 100 billion streaming events, corresponding to an integrated luminosity of 2.9~pb$^{-1}$.

\section{Heavy flavor reconstruction}

\begin{figure}
    \centering
    \includegraphics[width=0.32\linewidth]{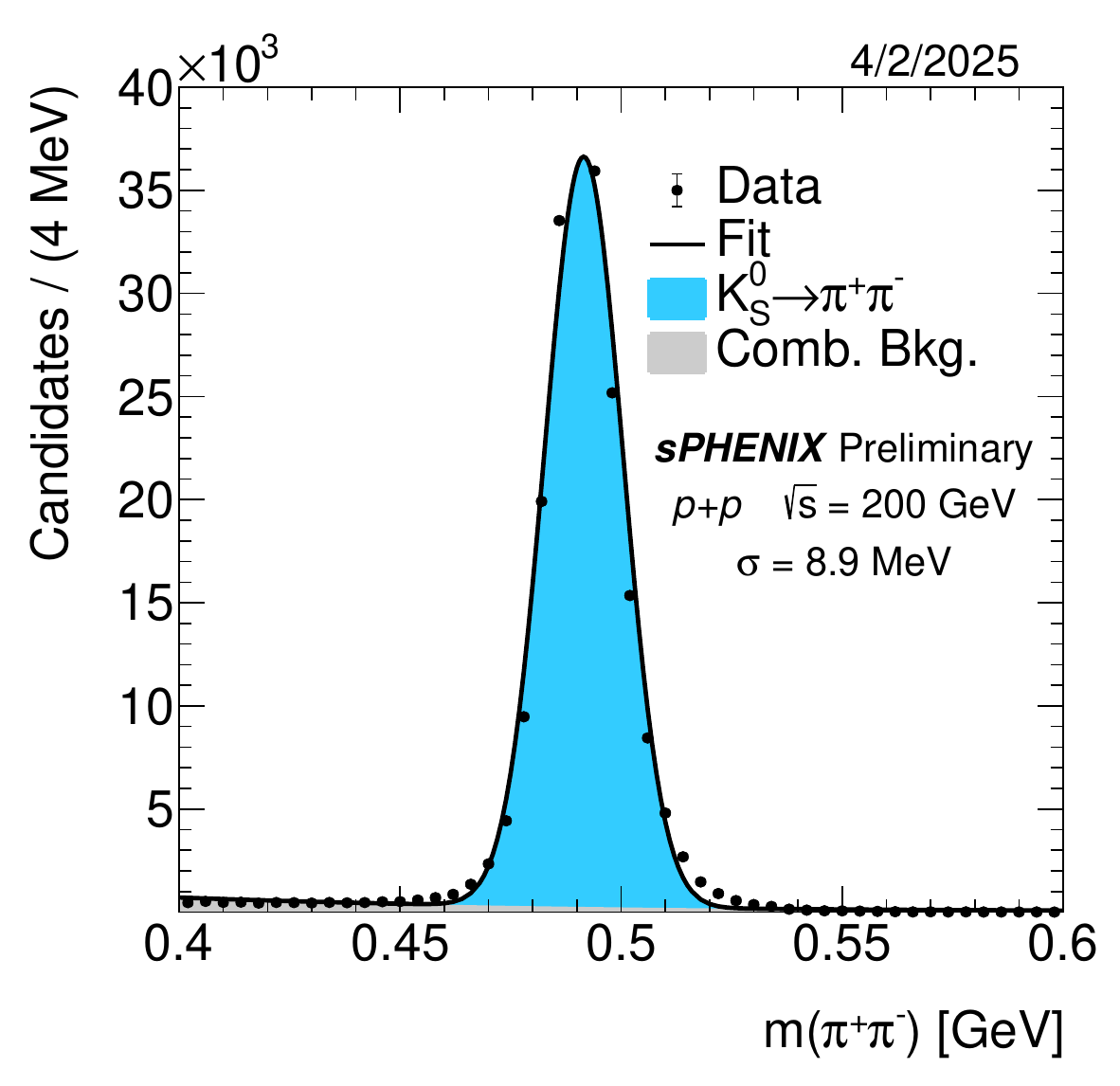}
    \includegraphics[width=0.32\linewidth]{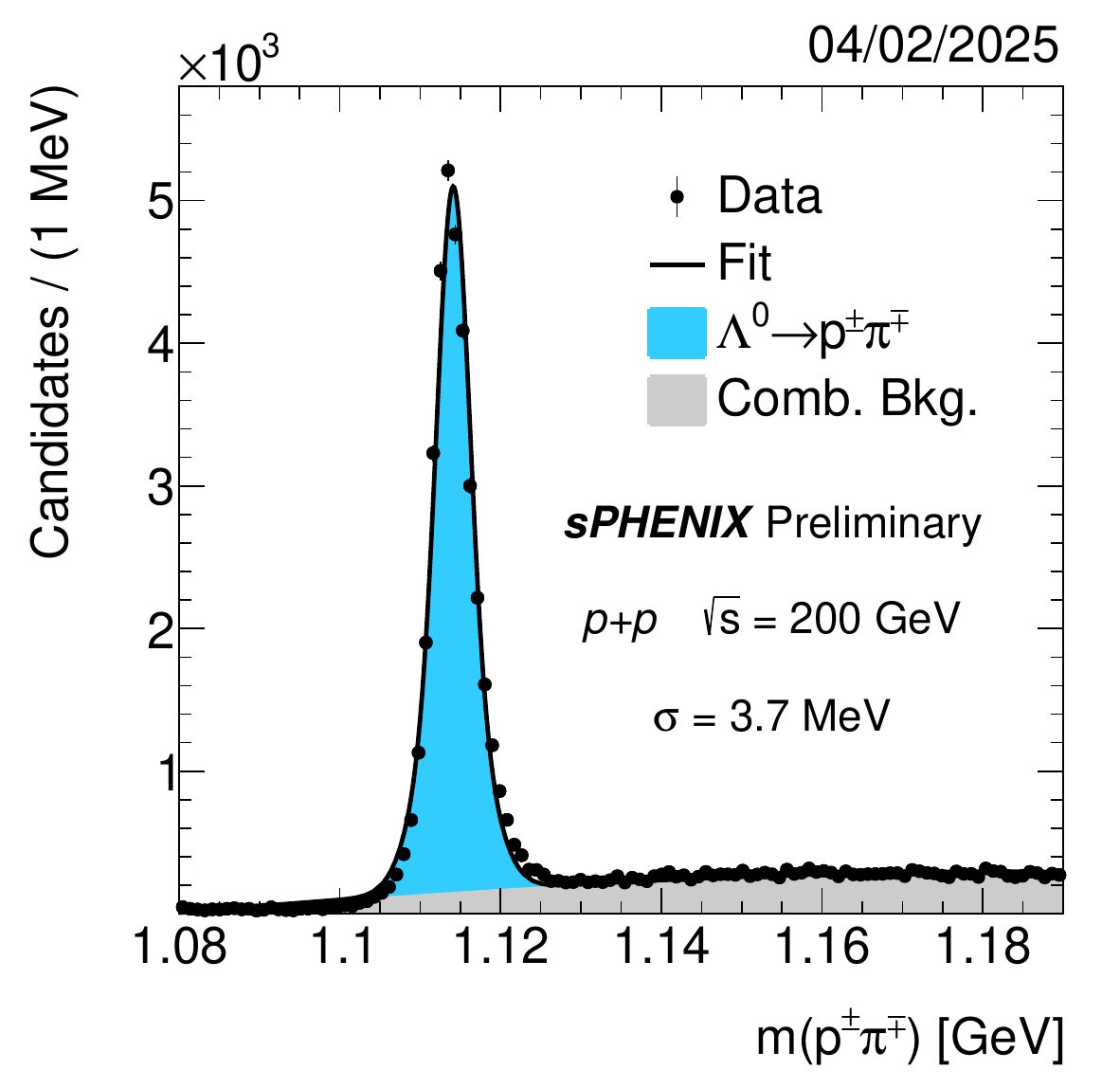}
    \\
    \includegraphics[width=0.32\linewidth]{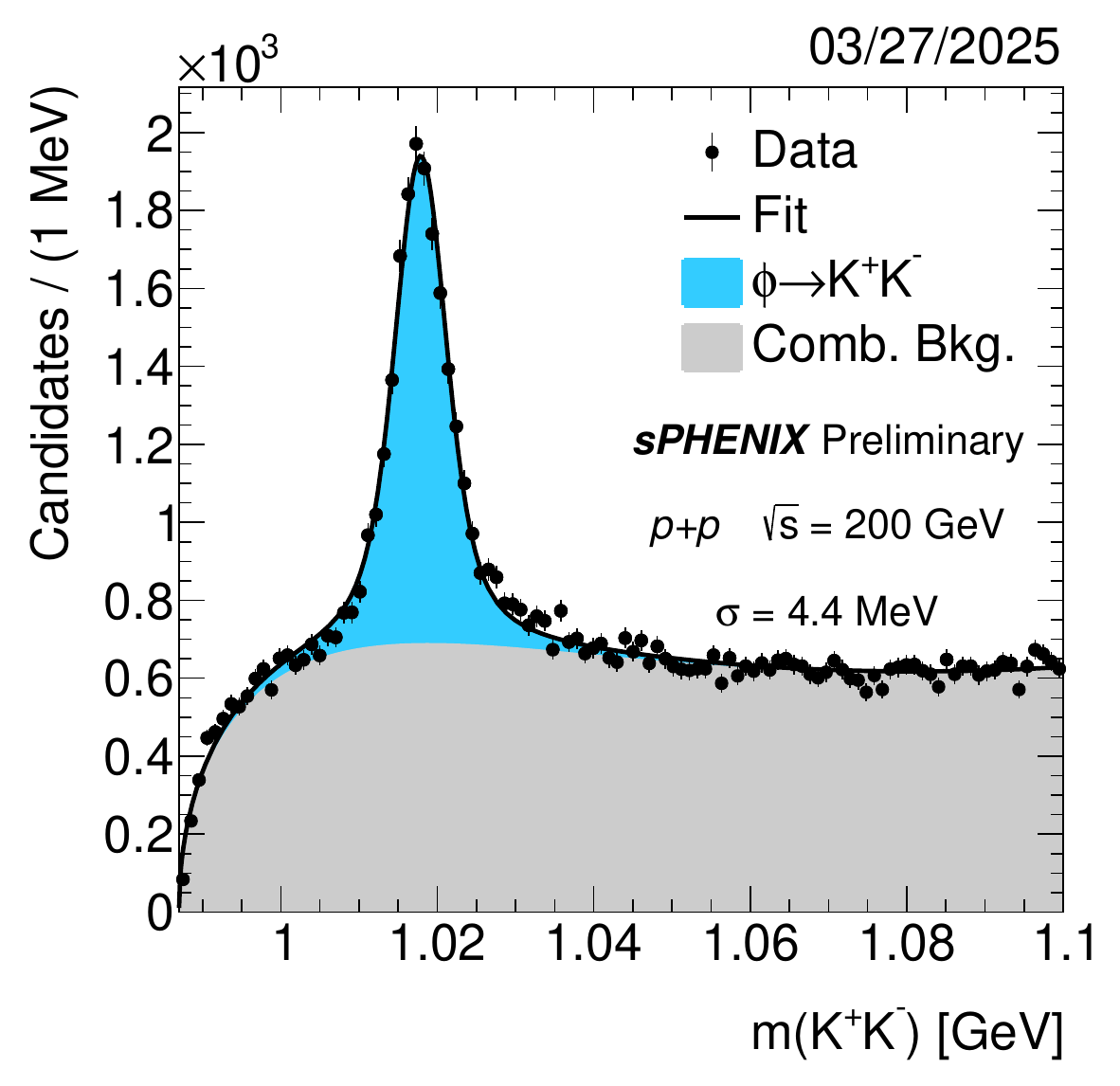}
    \includegraphics[width=0.32\linewidth]{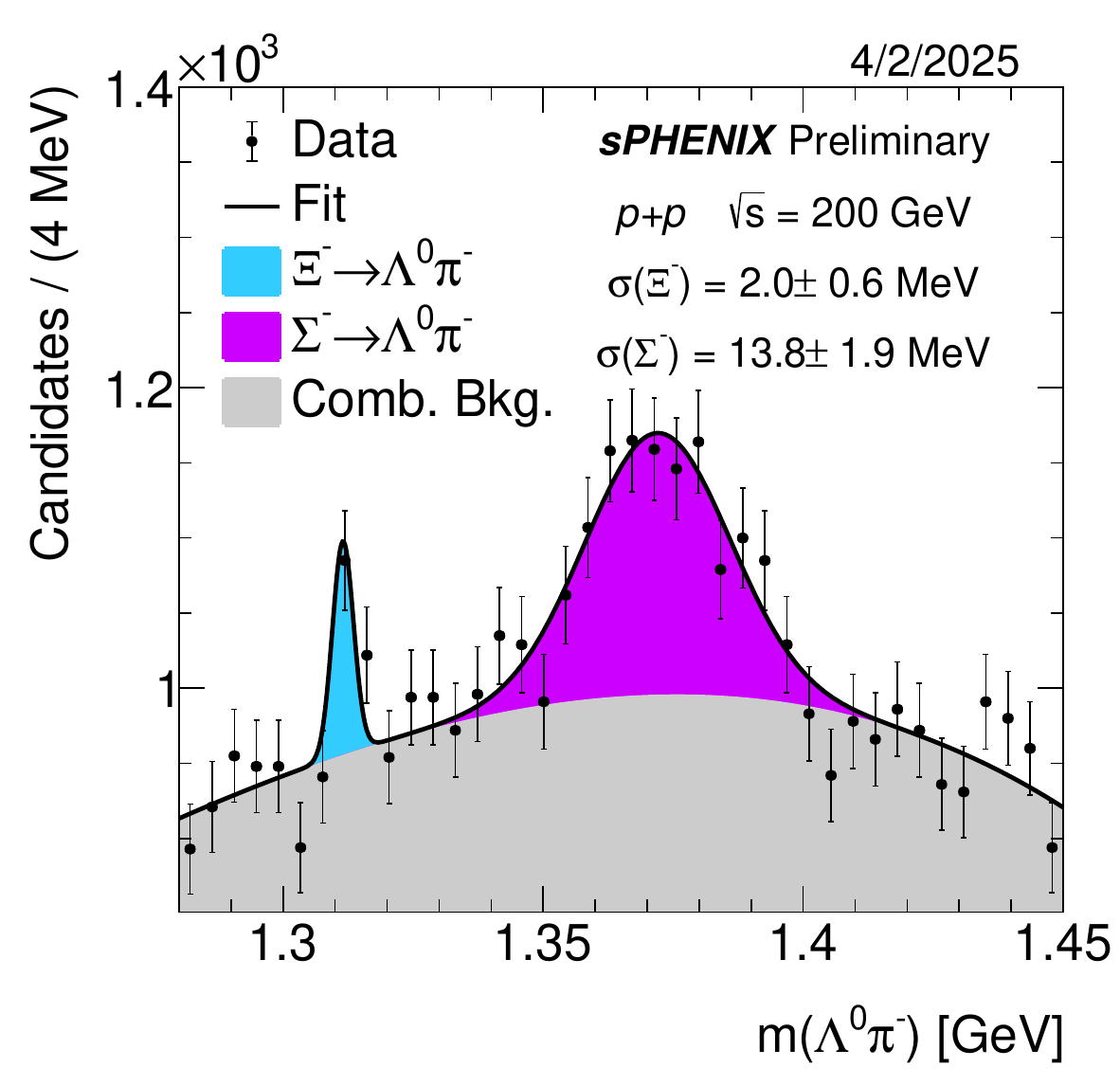}
    \caption{The invariant mass of reconstructed $\pi^+\pi^-$, $p\pi^-$, $K^+K^-$, and $\Lambda\pi^-$ combinations. The combinatorial background is shown in gray, and the signals of $K_S^0$, $\Lambda$, $\phi$, $\Xi^-$ and $\Sigma(1385)^-$ are shown in blue or purple.}
    \label{fig:lf}
\end{figure}

In the months following the completion of Run~24, the analysis has focused on a small subset of the tracking data, corresponding to roughly one hour of data taking. A major effort is underway to understand and calibrate the first full collision dataset recorded by the sPHENIX tracking system. This work includes the alignment of the silicon detectors using field-off data and the correction of TPC space-charge distortions with lamination fitting. After these initial calibrations, many well-known light-flavor resonances can already be reconstructed clearly. Examples include $K_S^0\to\pi^+\pi^-$, $\Lambda\to p\pi^-$, $\phi\to K^+K^-$, and more complex decay topologies such as $\Xi^-/\Sigma(1385)^-\to\Lambda\pi^-$ with $\Lambda\to p\pi^-$, as shown in Fig.~\ref{fig:lf}. These reconstructions are performed with the \textsc{KFParticle} package adapted to the sPHENIX software framework~\cite{kfparticlewiki}. They provide an important validation of the tracking and vertexing performance before moving to the more challenging open-heavy-flavor signals.

\begin{figure}[h]
    \centering
    \begin{subfigure}[b]{0.32\linewidth}
        \includegraphics[width=\linewidth]{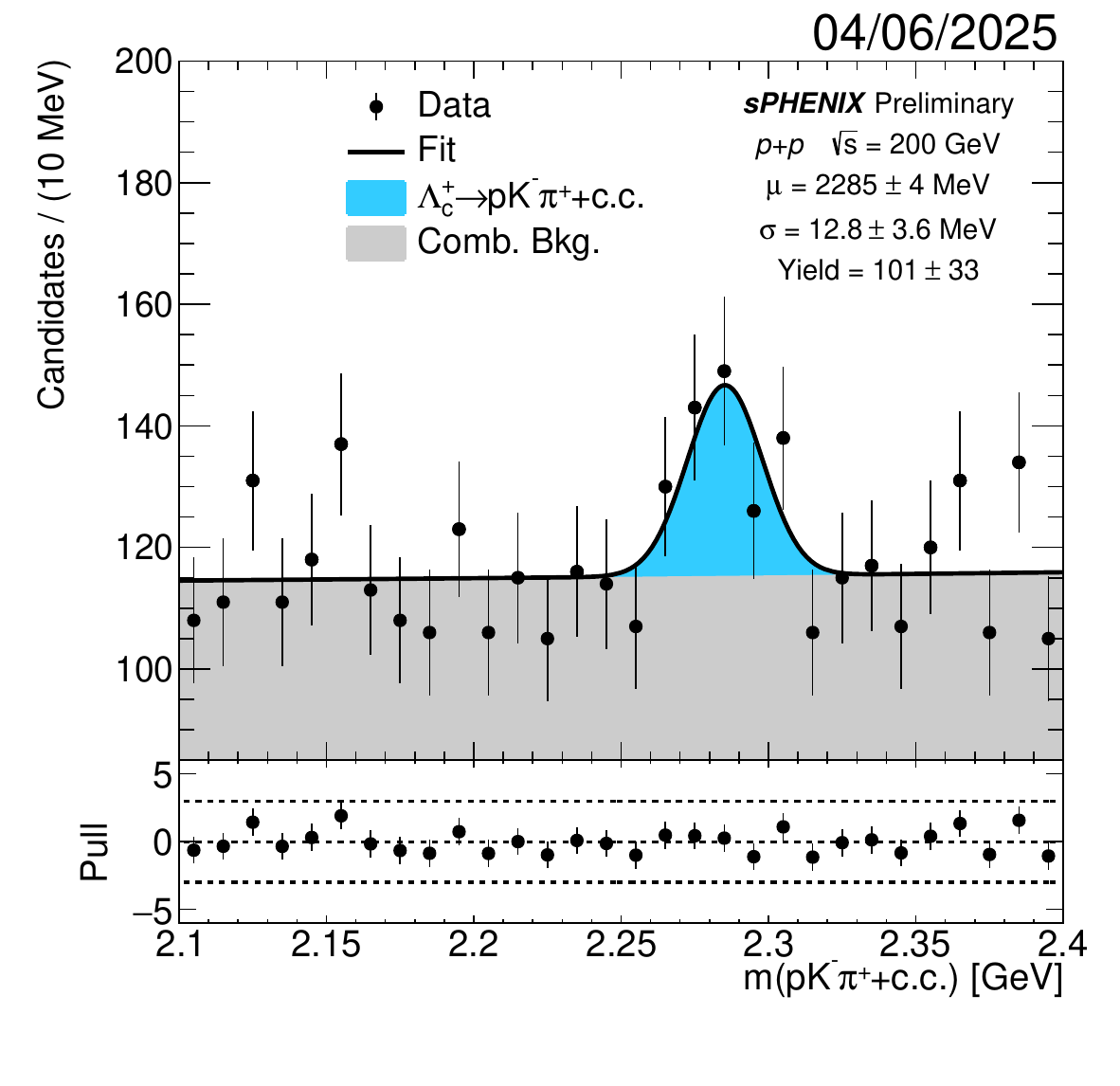}
        \caption{$\Lambda_c^+\to pK^-\pi^+$}
        \label{fig:hf-lc}
    \end{subfigure}
    \begin{subfigure}[b]{0.32\linewidth}
        \includegraphics[width=\linewidth]{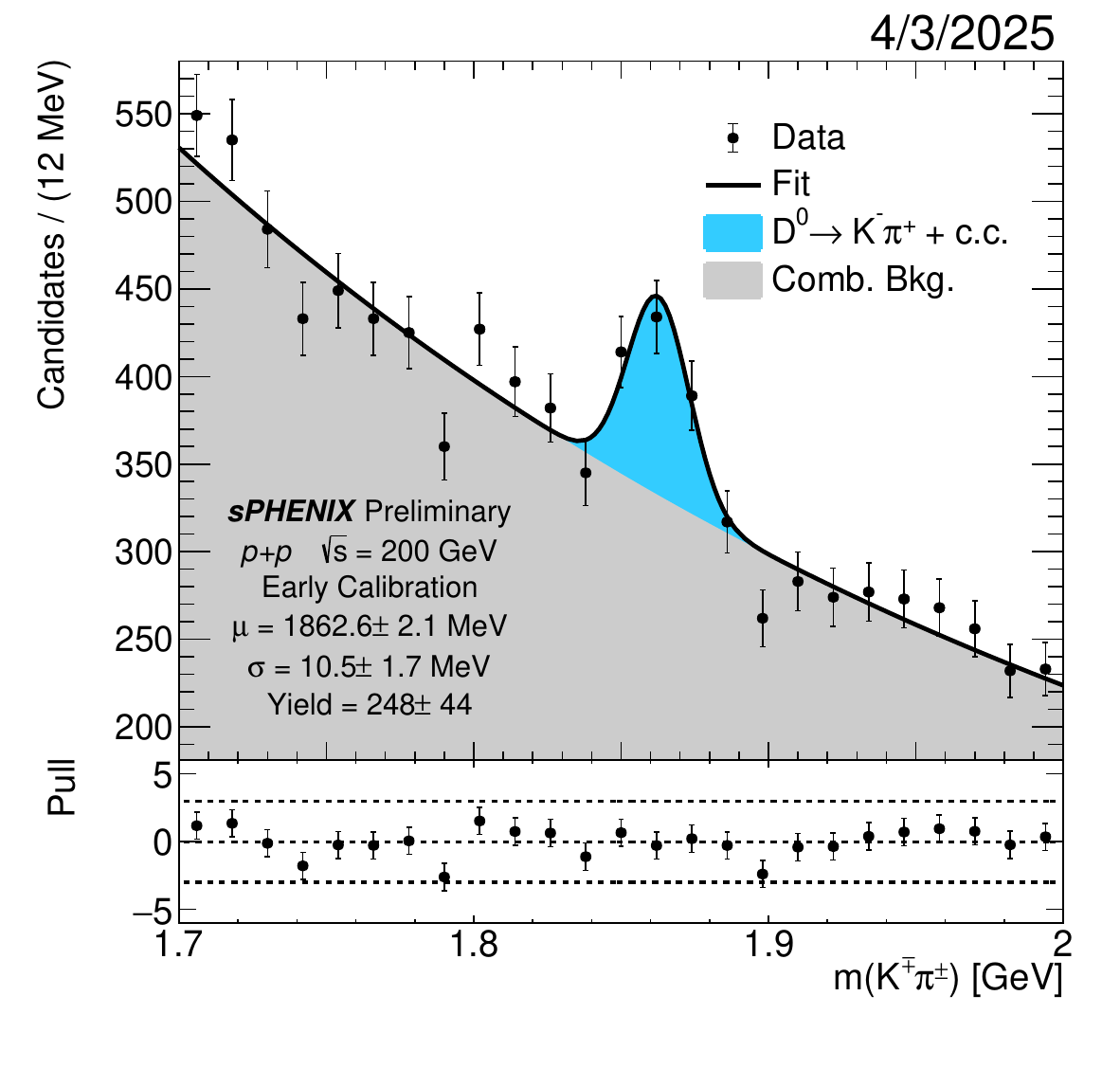}
        \caption{$D^0\to K^-\pi^+$}
        \label{fig:hf-d0}
    \end{subfigure}
    \\
    \begin{subfigure}[b]{0.32\linewidth}
        \includegraphics[width=\linewidth]{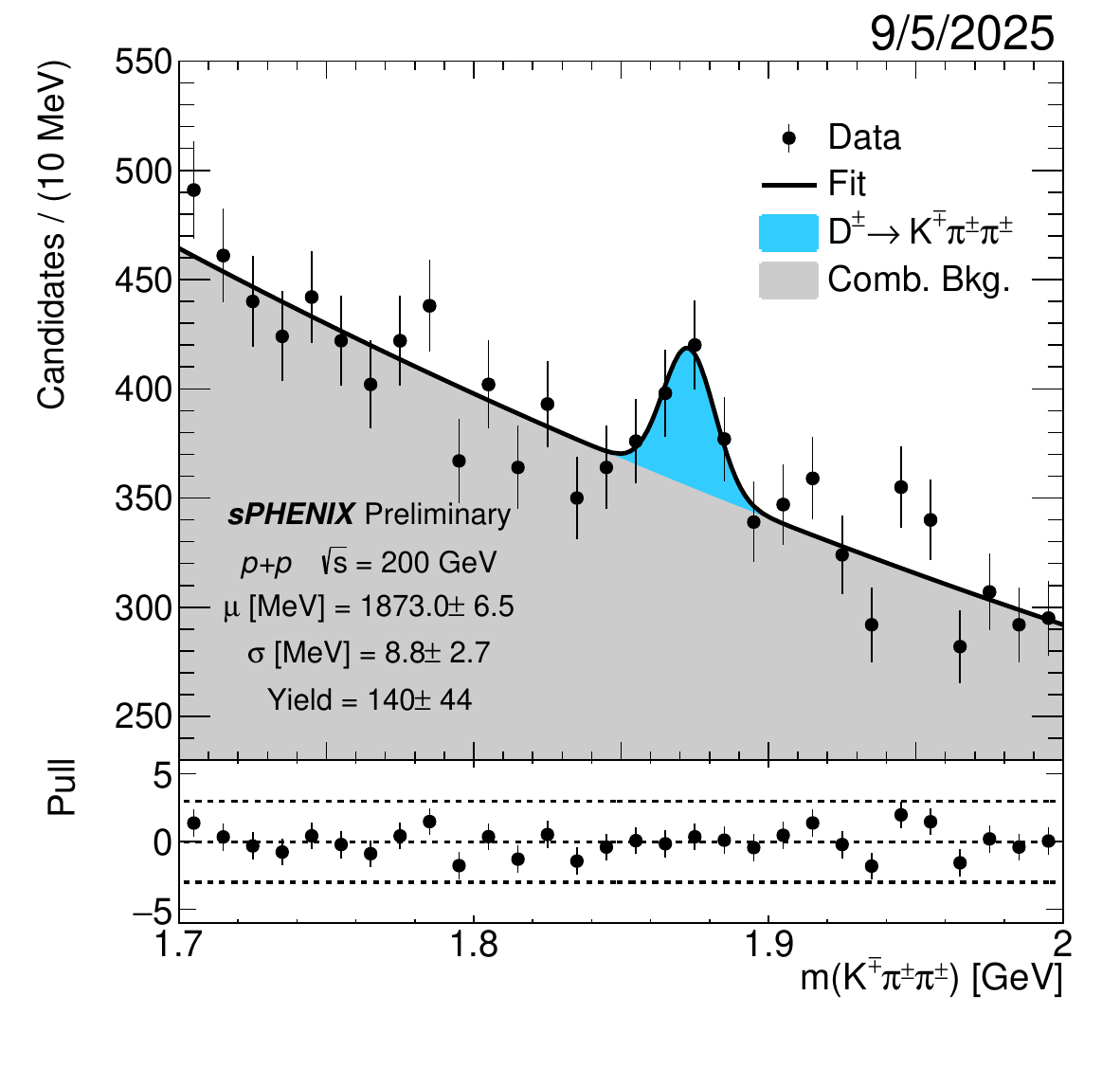}
        \caption{$D^+\to K^-\pi^+\pi^+$}
        \label{fig:hf-dp}
    \end{subfigure}
    \begin{subfigure}[b]{0.32\linewidth}
        \includegraphics[width=\linewidth]{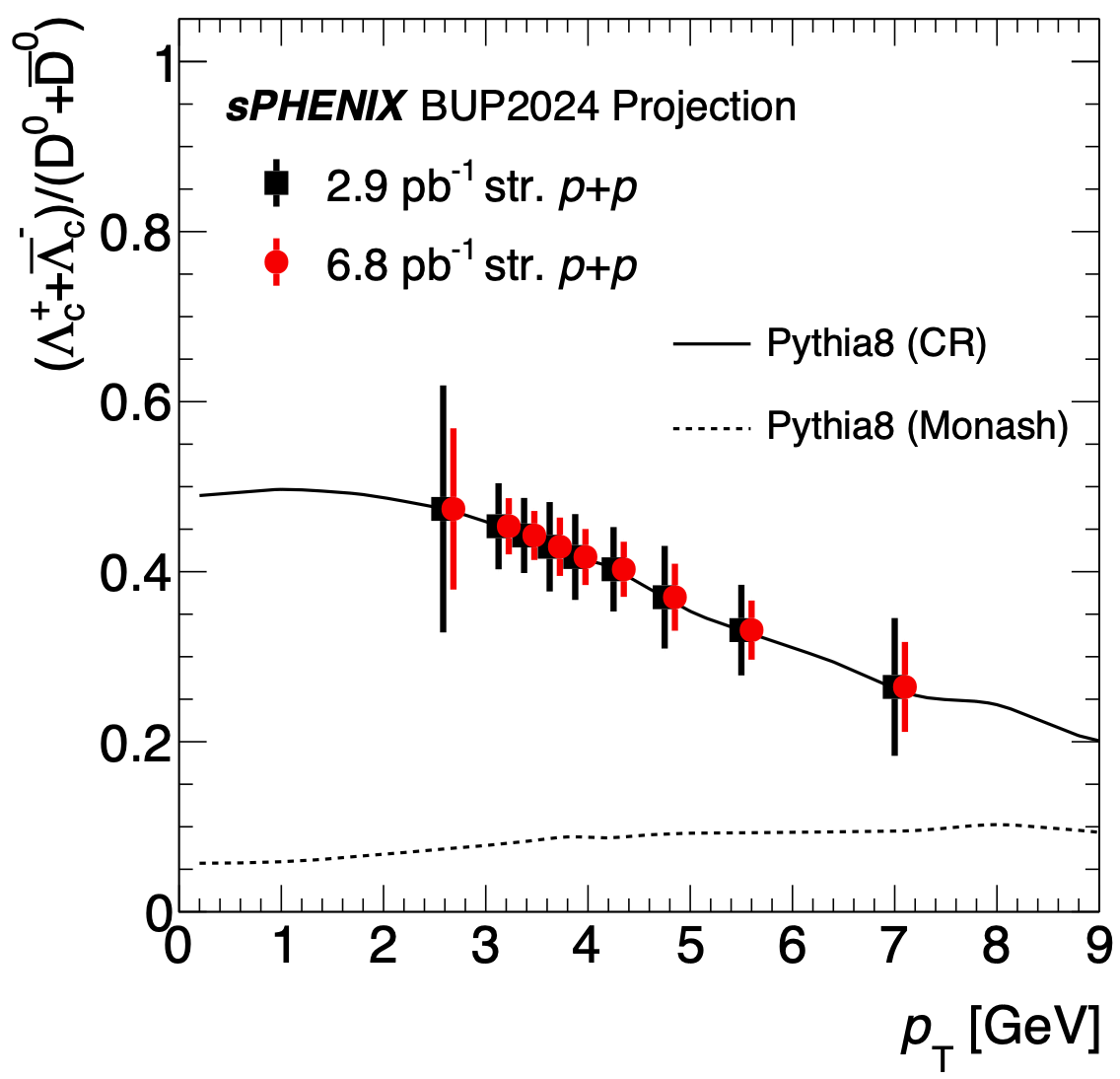}
        \caption{$\Lambda_c/D^0$}
        \label{fig:LcD0proj}
    \end{subfigure}
    \caption{(a) The invariant mass of $pK^-\pi^+$ combinations. (b) The invariant mass of the reconstructed $K^-\pi^+$ pairs. (c) The invariant mass of the reconstructed $K^-\pi^+\pi^+$ combinations. The combinatorial background is shown in gray, and the signals of $D^0, D^+$ mesons and $\Lambda_c^+$ baryon are shown in blue. (d) Statistical projection for the $\Lambda_c/D^0$ ratio as a function of $p_T$ in $p$+$p$ data.}
    \label{fig:hf}
\end{figure}

Compared with light-flavor resonances, open-charm hadrons have lower production yields, larger combinatorial backgrounds, and more demanding topological requirements, making their reconstruction substantially more challenging. Figure~\ref{fig:hf-lc} shows the first observation in $p$+$p$ collisions at RHIC of a $\Lambda_c^+$ signal reconstructed in the $pK^-\pi^+$ channel. Figure~\ref{fig:hf-d0} presents the first $D^0$ signal observed at sPHENIX through the $D^0\to K^-\pi^+$ decay, and Fig.~\ref{fig:hf-dp} shows the corresponding $D^+$ signal reconstructed in the $D^+\to K^-\pi^+\pi^+$ channel. Reconstruction of the $D_s^+$ meson is ongoing. Fig.~\ref{fig:LcD0proj} shows statistical projection for the $\Lambda_c/D^0$ ratio in $p$+$p$ data. The displaced secondary vertex and low-momentum $\mathrm{d}E/\mathrm{d}x$ information from the TPC are used to suppress the large combinatorial background. At present, the achievable track-pointing and momentum resolutions are still limited by the current understanding of the silicon alignment and TPC distortion corrections. Continued calibration improvements are expected to further enhance both the vertex and momentum resolutions, which are essential for precision measurements of the $\Lambda_c^+/D^0$ and $D_s^+/D^+$ yield ratios.

\section{Summary}
The sPHENIX experiment recorded 100 billion unbiased $p$+$p$ events at $\sqrt{s}=200$~GeV during Run~24. With its streaming readout capability and precision tracking system, the experiment has already demonstrated reconstruction of a broad set of resonances~\cite{sphenixpublic}, including the first $D^0$ signal from sPHENIX and the first observation of $\Lambda_c^+$ and $D^+$ in $p$+$p$ collisions at RHIC. These results establish the foundation for the first measurements of open-charm baryon-to-meson and strange-to-nonstrange charm-meson ratios at RHIC. In 2025, sPHENIX collected large datasets for Au+Au collisions with an integrated luminosity of 6.6~nb$^{-1}$, streaming $p$+$p$ collisions with 9.7~pb$^{-1}$, and streaming O+O collisions with 8.9~nb$^{-1}$. Together, these datasets will enable a broad heavy-flavor program and provide new opportunities to investigate charm production, hadronization, and the properties of the QGP.

\end{document}